\documentclass[conference]{IEEEtran}
\IEEEoverridecommandlockouts
\usepackage{cite}
\usepackage{amsmath,amssymb,amsfonts}
\usepackage{algorithm,algorithmic}
\usepackage{graphicx}
\usepackage{textcomp}
\def\BibTeX{{\rm B\kern-.05em{\sc i\kern-.025em b}\kern-.08em
		T\kern-.1667em\lower.7ex\hbox{E}\kern-.125emX}}

\usepackage{lipsum}
\usepackage{amsmath,amssymb,amsthm,mathrsfs,amsfonts,dsfont}

\DeclareMathOperator*{\argmin}{arg\,min}
\newcommand\norm[1]{\left\lVert#1\right\rVert}
\usepackage{tabularx}
\usepackage{subfigure}
\usepackage{xcolor}
\usepackage{epsfig}
\usepackage{epstopdf}
\usepackage{adjustbox}

\begin{document}
\title{Legitimate against Illegitimate IRSs on MISO Wiretap Channels\\
	\thanks{The work of Sepehr Rezvani and Eduard Jorswieck was supported in part by the German Research Foundation (DFG) under Grant JO 801/24-1. The authors acknowledge the financial support by the Federal Ministry of Education and Research of Germany in the program of "Souverän. Digital. Vernetzt." Joint project 6G-RIC, project identification number: 16KISK020K and 16KISK031.}
}

\author{
	\IEEEauthorblockN{Sepehr Rezvani, Pin-Hsun Lin, Martin Le, and Eduard Jorswieck}
	\IEEEauthorblockA{Institute for Communications Technology, Technische Universität Braunschweig, Braunschweig, Germany\\
	e-mails: \{Rezvani, Lin, Le, Jorswieck\}@ifn.ing.tu-bs.de}
}

\maketitle

\begin{abstract}
	The low-cost legitimate intelligent reflecting surfaces (IRSs) have been applied to the wiretap channel in physical layer security to enhance the secrecy rate. In practice, the eavesdropper can also deploy an IRS, namely illegitimate IRS, to deteriorate the secrecy rate. This paper studies the interplay between a transmitter, a legitimate IRS, and an illegitimate IRS in a multiple-input single-output (MISO) wiretap channel. We formulate a max-min secrecy rate problem, where the channel state and resource allocation information are available at the transmitter as well as the receivers. We aim to design an efficient transmit beamforming and phase shifting strategy of the legitimate IRS, under the worst-case secrecy rate achieved based on optimizing the phase shifting strategy of the illegitimate IRS. We propose three solution methods based on the gradient descent ascent (GDA), the alternate optimization (AO), and the mixed Nash equilibrium (NE) in zero-sum games in strategic form. Numerical results are provided to demonstrate the performance and convergence behavior of AO, GDA, and the mixed NE for continuous and discrete domains of IRSs' phase shifts.
\end{abstract}		


\section{Introduction}
\allowdisplaybreaks
\IEEEPARstart{I}{ntelligent} reflecting surface (IRS) has been developed as a key enabler to realize programmable and controllable signal propagation environment \cite{Renzo2019,9140329}. The IRS can be thought of as a low-cost (smart) thin metasurface including passive reflecting elements, each of which is capable of modifying the amplitude and phase of the electromagnetic waves by using external stimuli, resulting in higher spectral efficiency \cite{9140329,9343768}. It has been shown that the passive elements of IRSs lead to much less power consumption compared to the traditional active transceivers or relays \cite{Renzo2019,9140329,9343768}.

Recent research studies investigate the advantages of IRSs to improve the physical layer security (PLS) of wireless communications \cite{8743496,8723525,8847342,9146177,8972406,9133130,9201173,9446526,9187230,9262884}. The research studies on IRS-aided wiretap channels mainly focus on the advantages of legitimate IRSs under the control of the transmitter (Alice) to provide secure communications.
However, the eavesdropper (Eve) can also use low-cost IRSs, called illegitimate IRSs, to deteriorate the secrecy rate. There are few works on PLS with the existence of illegitimate IRSs. In \cite{Staat_Eve_IRS}, the authors consider a wiretap channel, where Eve uses an IRS to degrade the legitimate receiver's (Bob's) reception by a passive jamming. In \cite{Lyu_Eve_IRS}, a P2P channel with the existence of an IRS jammer is studied.
To the best of our knowledge, the interplay between legitimate and illegitimate IRSs on the secrecy rate is not yet studied in the literature.
In this work, we consider a new scenario, where both Bob and Eve use independent IRSs under the perfect channel state information\footnote{In our considered model, Eve needs to feedback the CSI to her own IRS's controller in order to efficiently tune her IRS's phase shifting elements for minimizing the secrecy rate. Hence, Eve is assumed to be active and detectable by Alice. The impact of imperfect CSI, and the case that Eve (and/or Eve's IRS) is undetectable are considered as future works.}. Moreover, the beamforming strategy (Alice's strategy) as well as the phase shifting strategy of Bob's IRS (Bob's strategy) are available at Eve when she is tuning her own IRSs phase shifting elements (Eve's strategy).
Our contributions are as follows:
\begin{itemize}
	\item We study the impacts of legitimate and illegitimate IRSs on the secrecy rate. In this scenario, we show that depending on some channel conditions, each IRS acts as a signal enhancer for its corresponding receiver or jammer for the other one.
	\item We design an efficient joint Alice's (beamforming) and Bob's (legitimate IRS) strategy to maximize the secrecy rate. To make the algorithm robust, we consider the worst-case secrecy rate achieved by optimizing Eve's (illegitimate IRS) strategy for any given Alice's and Bob's strategies. Hence, we formulate a novel max-min secrecy rate problem.
	\item We propose three solution methods based on gradient descent ascent (GDA), alternate optimization (AO), and non-cooperative game theory. We numerically evaluate the convergence behavior and performance of AO and GDA for continuous and discrete domains of IRSs' phase shifting elements.
\end{itemize}

\section{System Model and Problem Formulation}\label{Section sysmodel}
\subsection{System Model}
\begin{figure}
	\centering
	\includegraphics[scale=0.43]{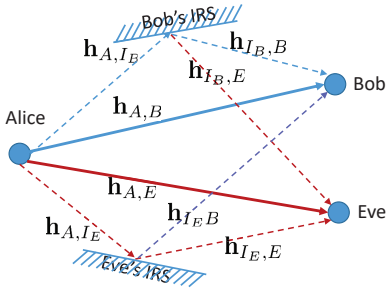}
	\caption{The exemplary model of an IRS-assisted MISO wiretap channel with the existence of an illegitimate IRS.}
	\label{Fig_sys}
\end{figure}

We consider the multiple-input single-output (MISO) wiretap channel, where a single transmitter (Alice) equipped with $M$ antennas communicates with a single-antenna legitimate user (Bob) in the presence of a single-antenna eavesdropper (Eve) overhearing the broadcast signal as shown in Fig. \ref{Fig_sys}.
Moreover, an IRS with $N_\text{B}$ elements under the control of Alice, namely legitimate IRS or Bob's IRS, is deployed. Bob's IRS is responsible for enhancing the data rate of Bob and/or degrading the data rate of Eve, thus enhancing the secrecy rate. Besides, Eve's IRS with $N_\text{E}$ reflecting elements is out of Alice's control, and is responsible for enhancing Eve's data rate and/or degrading Bob's data rate, thus degrading the secrecy rate. The sets of reflecting elements of Bob's and Eve's IRS are denoted by $\mathcal{N}_\text{B}=\{1,\dots,N_\text{B}\}$, and $\mathcal{N}_\text{E}=\{1,\dots,N_\text{E}\}$, respectively.
Theoretically, the reflection coefficient of each IRS element $n$ is modelled by $\alpha_n e^{j \beta_n}$, where $\alpha_n \in [0,1]$ and $\beta_n \in [0,2\pi)$ represent the amplitude and phase shift of this element, respectively \cite{9110912}. We assume the reflecting amplitude $\alpha_n=1$ for each reflecting element of Bob's and Eve's IRSs, i.e., each reflecting element $n$ can only tune the phase shift $\beta_n$. Thus the reflecting element is referred to as phase shifting element.
Let $\boldsymbol{\theta}_\text{B}=\left[\theta^\text{B}_1,\dots,\theta^\text{B}_{N_\text{B}}\right]^T$ and $\boldsymbol{\theta}_\text{E}=\left[\theta^\text{E}_1,\dots,\theta^\text{E}_{N_\text{E}}\right]^T$
denote the vectors of phase shifting coefficients of Bob's and Eve's IRS, respectively, where $\theta^\text{B}_m=e^{j\phi^\text{B}_m},~\forall m \in \mathcal{N}_\text{B}$, and $\theta^\text{E}_n=e^{j\phi^\text{E}_n},~\forall n \in \mathcal{N}_\text{E}$. The parameters $\phi^\text{B}_m  \in [0,2\pi)$ and $\phi^\text{E}_n  \in [0,2\pi)$ represent the phases of the $m$-th and $n$-th phase shifting elements of Bob's and Eve's IRS, respectively.
In practice, due to the hardware limitation, the phase shifts can take only a finite number of discrete values \cite{8930608}. Denoted by $L_\text{B}$ and $L_\text{E}$, the number of discrete values that each phase shifting element of Bob's and Eve's IRS can take, respectively. Without loss of generality, we assume that $\phi^\text{B}_m  \in \{\frac{2k\pi}{L_\text{B}} | k=0,\dots,(L_\text{B}-1)\},~\forall m \in \mathcal{N}_\text{B}$, and $\phi^\text{E}_n  \in \{\frac{2k\pi}{L_\text{E}} | k=0,\dots,(L_\text{E}-1)\},~\forall n \in \mathcal{N}_\text{E}$, \cite{9446526,9343768}.
Subsequently, we define the discrete set of possible phase shifting coefficients of each phase shifting element of Bob's and Eve's IRS, respectively, by $\mathcal{L}_\text{B}=\{ e^{j\frac{2k\pi}{L_\text{B}}} | k=0,\dots,(L_\text{B}-1)\}$, and $\mathcal{L}_\text{E}=\{ e^{j\frac{2k\pi}{L_\text{E}}} | k=0,\dots,(L_\text{E}-1)\}$.

In this system, Alice intends to send a confidential message by the independent and identically distributed Gaussian code symbol $x \in \mathbb{C}$ with zero mean and unit variance to Bob over a quasi-static flat-fading Gaussian wiretap channel. The beamforming vector is denoted by $\mathbf{w} \in \mathbb{C}^{M \times 1}$. The generally complex channel vector/matrix from Alice to Bob, Alice to Bob's IRS, Bob's IRS to Bob, Bob's IRS to Eve, Alice to Eve, Alice to Eve's IRS, Eve's IRS to Eve, and Eve's IRS to Bob are denoted by
$\mathbf{h}_\text{A,B} \in \mathbb{C}^{1 \times M}$,
$\mathbf{h}_\text{A,$\text{I}_\text{B}$} \in \mathbb{C}^{N_\text{B} \times M}$,
$\mathbf{h}_\text{$\text{I}_\text{B}$,B} \in \mathbb{C}^{1 \times N_\text{B}}$,
$\mathbf{h}_\text{$\text{I}_\text{B}$,E} \in \mathbb{C}^{1 \times N_\text{B}}$,
$\mathbf{h}_\text{A,E} \in \mathbb{C}^{1 \times M}$,
$\mathbf{h}_\text{A,$\text{I}_\text{E}$} \in \mathbb{C}^{N_\text{E} \times M}$,
$\mathbf{h}_\text{$\text{I}_\text{E}$,E} \in \mathbb{C}^{1 \times N_\text{E}}$,
$\mathbf{h}_\text{$\text{I}_\text{E}$,B} \in \mathbb{C}^{1 \times N_\text{E}}$,
respectively. We assume that the perfect CSI of all the links is available at all the nodes. The received signal at Bob and Eve can thus be formulated, respectively by\footnote{Due to the "double fading" effect, the powers reflected by IRSs two or more times are much smaller than those of signals reflected one time, thus the double reflection effect is ignored in this paper.}
\begin{align}
	y_B &= \underbrace{\left( \mathbf{h}_\text{A,B} + \mathbf{h}_\text{$\text{I}_\text{B}$,B} \mathbf{\Theta}_\text{B} \mathbf{h}_\text{A,$\text{I}_\text{B}$} + \mathbf{h}_\text{$\text{I}_\text{E}$,B} \mathbf{\Theta}_\text{E} \mathbf{h}_\text{A,$\text{I}_\text{E}$} \right)}_{\mathbf{h}_\text{B} \left(\mathbf{\Theta}_\text{B},\mathbf{\Theta}_\text{E}\right)} \mathbf{w} x + n_B,\label{signal Bob}\\
	y_E &= \underbrace{\left( \mathbf{h}_\text{A,E} + \mathbf{h}_\text{$\text{I}_\text{E}$,E} \mathbf{\Theta}_\text{E} \mathbf{h}_\text{A,$\text{I}_\text{E}$} + \mathbf{h}_\text{$\text{I}_\text{B}$,E} \mathbf{\Theta}_\text{B} \mathbf{h}_\text{A,$\text{I}_\text{B}$} \right)}_{\mathbf{h}_\text{E}\left(\mathbf{\Theta}_\text{B},\mathbf{\Theta}_\text{E}\right)} \mathbf{w} x + n_E,\label{signal Eve}
\end{align}
in which $\mathbf{\Theta}_\text{B} = \text{diag}\left(\boldsymbol{\theta}_\text{B}\right)$, $\mathbf{\Theta}_\text{E} = \text{diag}\left(\boldsymbol{\theta}_\text{E}\right)$, and $n_B$ and $n_E$ are the independent zero-mean additive white Gaussian noises (AWGNs) at Bob and Eve with variances $\sigma^2_B$ and $\sigma^2_E$, respectively. The row vectors $\mathbf{h}_\text{B}\left(\mathbf{\Theta}_\text{B},\mathbf{\Theta}_\text{E}\right) \in \mathbb{C}^{1 \times M}$, and $\mathbf{h}_\text{E}\left(\mathbf{\Theta}_\text{B},\mathbf{\Theta}_\text{E}\right) \in \mathbb{C}^{1 \times M}$ denote the effective/equivalent channel gains between Alice and Bob, and between Alice and Eve, respectively.
The secrecy capacity is thus given by\footnote{The second term in \eqref{secrecy rate} is to deteriorate the secrecy capacity. For convenience, we call it "Eve's rate" although Eve cannot decode the received information.}
\begin{multline}\label{secrecy rate}
	C_\text{s} (\mathbf{w},\mathbf{\Theta}_\text{B},\mathbf{\Theta}_\text{E})=
	\underbrace{\log_2 \left( 1 + \frac{ \lvert \mathbf{h}_\text{B}\left(\mathbf{\Theta}_\text{B},\mathbf{\Theta}_\text{E}\right) \mathbf{w} \rvert^2}{\sigma^2_B} \right)}_{\text{Bob's capacity}}
	\\
	-\underbrace{\log_2 \left( 1 + \frac{ \lvert \mathbf{h}_\text{E}\left(\mathbf{\Theta}_\text{B},\mathbf{\Theta}_\text{E}\right) \mathbf{w} \rvert^2}{\sigma^2_E} \right)}_{\text{Eve's rate}}.
\end{multline}

\subsection{Problem Formulation}
In this system, Alice intends to optimize the beamforming and legitimate IRS's phase shifting strategies to maximize the secrecy rate. We consider the worst-case scenario, where Eve can access the information about the adopted\footnote{We assume that the optimized control bits (Bob's strategy) sent from Alice to Bob's IRS over the wireless link can be eavesdropped by Eve.} $\mathbf{w}$, and $\mathbf{\Theta}_\text{B}$, and then optimize $\mathbf{\Theta}_\text{E}$.
The secrecy capacity in the worst-case scenario is given by $\min\limits_{\mathbf{\Theta}_\text{E}}~ C_\text{s} (\mathbf{w},\mathbf{\Theta}_\text{B},\mathbf{\Theta}_\text{E})$. In this work, we aim at designing efficient joint active ($\mathbf{w}$) and passive ($\mathbf{\Theta}_\text{B}$) beamforming strategies to maximize the worst-case secrecy capacity. The max-min secrecy capacity problem is formulated by
\begin{subequations}\label{main problem}
	\begin{align}\label{obf main}
		\max\limits_{\mathbf{w},\mathbf{\Theta}_\text{B}}~\hspace{.0 cm}
		&\min\limits_{\mathbf{\Theta}_\text{E}} C_\text{s} (\mathbf{w},\mathbf{\Theta}_\text{B},\mathbf{\Theta}_\text{E})
		\\
		\text{s.t.}~~
		\label{constraint power}
		& \norm{\mathbf{w}}^2_2 \leq P,
		\\
		\label{constraint theta}
		& \theta^\text{B}_m \in \mathcal{L}_\text{B},~\forall m \in \mathcal{N}_\text{B},
		\\
		\label{constraint phi}
		& \theta^\text{E}_n \in \mathcal{L}_\text{E},~\forall n \in \mathcal{N}_\text{E},
	\end{align}
\end{subequations}
where $P$ denotes the maximum available transmit power of Alice. In \eqref{obf main}, we omit the operator $\{.\}^+$ without loss of optimality, since the optimal value is always non-negative\footnote{For the case that the optimal value is negative, Alice does not send any data to Bob, and subsequently, the secrecy capacity will be zero.}.

\section{Solution Algorithms}
The optimization problem \eqref{main problem} is nonconvex, due to the nonconcavity of the objective function \eqref{obf main} with respect to either $\mathbf{w}$ or $(\mathbf{\Theta}_\text{B},\mathbf{\Theta}_\text{E})$, and nonconvexity of constraints \eqref{constraint theta} and \eqref{constraint phi} with discrete domains. In this way, it is still difficult to obtain the globally optimal solution of \eqref{main problem}. In the following, we propose efficient suboptimal solution methods.

\subsection{Alternate Optimization}\label{Sec_AO_SDR}
To make problem \eqref{main problem} more tractable, we propose a three-step AO method as follows: 1) Finding ${\Theta}_\text{B}$ for the given $\left(\mathbf{\Theta}_\text{E},\mathbf{w}\right)$; 2) Finding $\mathbf{w}$ for the given $\left(\mathbf{\Theta}_\text{B},\mathbf{\Theta}_\text{E}\right)$; 3) Finding $\mathbf{\Theta}_\text{E}$ for the given $\left(\mathbf{\Theta}_\text{B},\mathbf{w}\right)$. Note that we should optimize $\mathbf{\Theta}_\text{E}$ at the last step of each AO iteration, due to considering the worst-case secrecy rate.

\subsubsection{Finding $\mathbf{\Theta}_\text{B}$}\label{subsec find thetaB}
For any given $\left(\mathbf{w},\mathbf{\Theta}_\text{E}\right)$, the main problem \eqref{main problem} can be equivalently transformed to the following maximization problem as
\begin{align}\label{problem 1a}
	\max\limits_{\mathbf{\Theta}_\text{B}}~
	\frac{\frac{1}{\sigma^2_B} \lvert \mathbf{h}_\text{B}\left(\mathbf{\Theta}_\text{B},\mathbf{\Theta}_\text{E}\right) \mathbf{w} \rvert^2 + 1}
	{\frac{1}{\sigma^2_E} \lvert \mathbf{h}_\text{E}\left(\mathbf{\Theta}_\text{B},\mathbf{\Theta}_\text{E}\right) \mathbf{w} \rvert^2 + 1}~~~~
	\text{s.t.}~\eqref{constraint theta}.
\end{align}
To make problem \eqref{problem 1a} tractable, we first relax \eqref{constraint theta}, by letting each element in $\mathbf{\Theta}_\text{B}$ to be continuous. In this way, we replace \eqref{constraint theta} with the unit-norm constraint $|\theta^\text{B}_m| = 1,~\forall m \in \mathcal{N}_\text{B}$.
According to $$\mathbf{h}_\text{$\text{I}_\text{B}$,B} \mathbf{\Theta}_\text{B} \mathbf{h}_\text{A,$\text{I}_\text{B}$} = \boldsymbol{\theta}^T_\text{B} \text{diag} \left( \mathbf{h}_\text{$\text{I}_\text{B}$,B} \right) \mathbf{h}_\text{A,$\text{I}_\text{B}$},$$
and
$$\mathbf{h}_\text{$\text{I}_\text{B}$,E} \mathbf{\Theta}_\text{B} \mathbf{h}_\text{A,$\text{I}_\text{B}$} = \boldsymbol{\theta}^T_\text{B} \text{diag} \left( \mathbf{h}_\text{$\text{I}_\text{B}$,E} \right) \mathbf{h}_\text{A,$\text{I}_\text{B}$},$$
we have \cite{8723525}
\begin{align}\label{equiv 1}
	\frac{1}{\sigma^2_B} \lvert \mathbf{h}_\text{B}\left(\mathbf{\Theta}_\text{B},\mathbf{\Theta}_\text{E}\right) \mathbf{w} \rvert^2 = \boldsymbol{\bar{\theta}}^H_\text{B} \mathbf{\bar{H}}_\text{B} \left(\mathbf{\Theta}_\text{E}\right) \boldsymbol{\bar{\theta}}_\text{B} + \mathbf{\bar{h}}_\text{B} \left(\mathbf{\Theta}_\text{E}\right),
\end{align}
and
\begin{align}\label{equiv 2}
	\frac{1}{\sigma^2_E} \lvert \mathbf{h}_\text{E}\left(\mathbf{\Theta}_\text{B},\mathbf{\Theta}_\text{E}\right) \mathbf{w} \rvert^2 = \boldsymbol{\bar{\theta}}^H_\text{B} \mathbf{\bar{H}}_\text{E} \left(\mathbf{\Theta}_\text{E}\right) \boldsymbol{\bar{\theta}}_\text{B} + \mathbf{\bar{h}}_\text{E} \left(\mathbf{\Theta}_\text{E}\right),
\end{align}
where $\boldsymbol{\bar{\theta}}_\text{B} = \left[ \boldsymbol{\theta}^T_\text{B} , 1 \right]^T$, 
$$
\mathbf{\bar{h}}_\text{B} = \frac{\left( \mathbf{h}_\text{A,B} + \mathbf{h}_\text{$\text{I}_\text{E}$,B} \mathbf{\Theta}_\text{E} \mathbf{h}_\text{A,$\text{I}_\text{E}$} \right)^* \mathbf{w}^* \mathbf{w}^T \left( \mathbf{h}_\text{A,B} + \mathbf{h}_\text{$\text{I}_\text{E}$,B} \mathbf{\Theta}_\text{E} \mathbf{h}_\text{A,$\text{I}_\text{E}$} \right)^T}{\sigma^2_B},
$$
$$
\mathbf{\bar{h}}_\text{E} = \frac{\left( \mathbf{h}_\text{A,E} + \mathbf{h}_\text{$\text{I}_\text{E}$,E} \mathbf{\Theta}_\text{E} \mathbf{h}_\text{A,$\text{I}_\text{E}$} \right)^* \mathbf{w}^* \mathbf{w}^T \left( \mathbf{h}_\text{A,E} + \mathbf{h}_\text{$\text{I}_\text{E}$,E} \mathbf{\Theta}_\text{E} \mathbf{h}_\text{A,$\text{I}_\text{E}$} \right)^T}{\sigma^2_E},
$$
and $\mathbf{\bar{H}}_\text{B}  \left(\mathbf{\Theta}_\text{E}\right)$ and $\mathbf{\bar{H}}_\text{E}  \left(\mathbf{\Theta}_\text{E}\right)$ are given by \eqref{Hbar B} and \eqref{Hbar E}, respectively.
\begin{figure*}[!t]
	\begin{align}\label{Hbar B}
		&\mathbf{\bar{H}}_\text{B}  \left(\mathbf{\Theta}_\text{E}\right)= \frac{1}{\sigma^2_B} \begin{bmatrix} \text{diag} \left( \mathbf{h}_\text{$\text{I}_\text{B}$,B} \right) \mathbf{h}_\text{A,$\text{I}_\text{B}$} \mathbf{w} \mathbf{w}^H \mathbf{h}^H_\text{A,$\text{I}_\text{B}$} \text{diag} \left( \mathbf{h}_\text{$\text{I}_\text{B}$,B} \right) & \text{diag} \left( \mathbf{h}_\text{$\text{I}_\text{B}$,B} \right) \mathbf{h}_\text{A,$\text{I}_\text{B}$} \mathbf{w} \mathbf{w}^H \left( \mathbf{h}^H_\text{A,B} + \mathbf{h}^H_\text{A,$\text{I}_\text{E}$} \mathbf{\Theta}_\text{E}^H \mathbf{h}^H_\text{$\text{I}_\text{E}$,B} \right) \\
			\left( \mathbf{h}_\text{A,B} + \mathbf{h}_\text{$\text{I}_\text{E}$,B} \mathbf{\Theta}_\text{E} \mathbf{h}_\text{A,$\text{I}_\text{E}$} \right) \mathbf{w} \mathbf{w}^H \mathbf{h}_\text{A,$\text{I}_\text{B}$}^H\text{diag} \left( \mathbf{h}_\text{$\text{I}_\text{B}$,B}^H \right) & 0 \end{bmatrix},\\
		&\mathbf{\bar{H}}_\text{E}  \left(\mathbf{\Theta}_\text{E}\right)= \frac{1}{\sigma^2_E} \begin{bmatrix} \text{diag} \left( \mathbf{h}_\text{$\text{I}_\text{B}$,E} \right) \mathbf{h}_\text{A,$\text{I}_\text{B}$} \mathbf{w} \mathbf{w}^H \mathbf{h}^H_\text{A,$\text{I}_\text{B}$} \text{diag} \left( \mathbf{h}_\text{$\text{I}_\text{B}$,E} \right) & \text{diag} \left( \mathbf{h}_\text{$\text{I}_\text{B}$,E} \right) \mathbf{h}_\text{A,$\text{I}_\text{B}$} \mathbf{w} \mathbf{w}^H \left( \mathbf{h}^H_\text{A,E} + \mathbf{h}^H_\text{A,$\text{I}_\text{E}$} \mathbf{\Theta}_\text{E}^H \mathbf{h}^H_\text{$\text{I}_\text{E}$,E} \right) \\
			\left( \mathbf{h}_\text{A,E} + \mathbf{h}_\text{$\text{I}_\text{E}$,E} \mathbf{\Theta}_\text{E} \mathbf{h}_\text{A,$\text{I}_\text{E}$} \right) \mathbf{w} \mathbf{w}^H \mathbf{h}_\text{A,$\text{I}_\text{B}$}^H\text{diag} \left( \mathbf{h}_\text{$\text{I}_\text{B}$,E}^H \right) & 0 \end{bmatrix},\label{Hbar E}
	\end{align}
\end{figure*}
Here, the superscript $^*$ denotes the complex conjugate operation.
According to \eqref{equiv 1} and \eqref{equiv 2}, the relaxed form of \eqref{problem 1a} can be rewritten as
\begin{subequations}\label{problem theta 2} 
	\begin{align}\label{obf problem theta 2}
		\max\limits_{\boldsymbol{\bar{\theta}}_\text{B}}~
		& \frac{\boldsymbol{\bar{\theta}}^H_\text{B} \mathbf{\bar{H}}_\text{B} \left(\mathbf{\Theta}_\text{E}\right) \boldsymbol{\bar{\theta}}_\text{B} + \mathbf{\bar{h}}_\text{B} \left(\mathbf{\Theta}_\text{E}\right) + 1}
		{\boldsymbol{\bar{\theta}}^H_\text{B} \mathbf{\bar{H}}_\text{E} \left(\mathbf{\Theta}_\text{E}\right) \boldsymbol{\bar{\theta}}_\text{B} + \mathbf{\bar{h}}_\text{E} \left(\mathbf{\Theta}_\text{E}\right) + 1}
		\\
		\text{s.t.}~~
		\label{continuous thetaE}
		& \boldsymbol{\bar{\theta}}^H_\text{B} \mathbf{E}_m \boldsymbol{\bar{\theta}}_\text{B}=1,~\forall m \in \mathcal{N}_\text{B},
	\end{align}
\end{subequations}
where $\mathbf{E}_m$ is an $(N_\text{B}+1) \times (N_\text{B}+1)$ diagonal matrix, whose $(i, j)$-th element, represented by $[\mathbf{E}_m]_{(i,j)}$, is 
\begin{equation}\label{rate2brac}
	[\mathbf{E}_m]_{(i,j)}=
	\left\{
	\begin{array}{ll}
		1, &
		\hbox{if $i = j = m$;} \\
		0, & \hbox{o.w.}\
	\end{array}
	\right.
\end{equation}
The quadratic equality constraints in \eqref{continuous thetaE} are nonconvex. To this end, we apply the well-known semidefinite relaxation (SDR) technique. In this approach, we define $\boldsymbol{\bar{\Theta}}_\text{B} = \boldsymbol{\bar{\theta}}_\text{B} \boldsymbol{\bar{\theta}}^H_\text{B}$ and formulate the relaxed form of \eqref{problem theta 2} (without the rank-1 constraint $\text{rank}(\boldsymbol{\bar{\Theta}}_\text{B})=1$) as follows:
\begin{subequations}\label{problem theta 3} 
	\begin{align}\label{obf problem theta 3}
		\max\limits_{\boldsymbol{\bar{\Theta}}_\text{B} \succcurlyeq 0}~
		& \frac{\text{tr}\left(\mathbf{\bar{H}}_\text{B} \left(\mathbf{\Theta}_\text{E}\right) \boldsymbol{\bar{\Theta}}_\text{B}\right) + \mathbf{\bar{h}}_\text{B}\left(\mathbf{\Theta}_\text{E}\right) + 1}
		{\text{tr}\left(\mathbf{\bar{H}}_\text{E}\left(\mathbf{\Theta}_\text{E}\right)  \boldsymbol{\bar{\Theta}}_\text{B}\right) + \mathbf{\bar{h}}_\text{E}\left(\mathbf{\Theta}_\text{E}\right) + 1}
		\\
		\text{s.t.}~~
		\label{continuous thetaE relax}
		& \text{tr}\left(\mathbf{E}_m \boldsymbol{\bar{\Theta}}_\text{B}\right)=1,~\forall m \in \mathcal{N}_\text{B}.
	\end{align}
\end{subequations}
According to the Charnes-Cooper transformation (CCT) \cite{6728676}, we define  
$\lambda = \left(\text{tr}\left(\mathbf{\bar{H}}_\text{E}\left(\mathbf{\Theta}_\text{E}\right)  \boldsymbol{\bar{\Theta}}_\text{B}\right) + \mathbf{\bar{h}}_\text{E}\left(\mathbf{\Theta}_\text{E}\right) + 1\right)^{-1}$, and $\mathbf{\tilde{\Theta}}_\text{B} = \lambda \boldsymbol{\bar{\Theta}}_\text{B}$. In this way, \eqref{problem theta 3} can be rewritten as
\begin{subequations}\label{problem theta 4}
\begin{align}\label{obf problem theta 4}
	\hspace{-0.3cm}\max\limits_{\mathbf{\tilde{\Theta}}_\text{B} ,\lambda }~
	&\text{tr}\left(\mathbf{\bar{H}}_\text{B} \left(\mathbf{\Theta}_\text{E}\right)  \mathbf{\tilde{\Theta}}_\text{B}\right) + \lambda \left(\mathbf{\bar{h}}_\text{B} \left(\mathbf{\Theta}_\text{E}\right) + 1\right)
	\\
	\text{s.t.}~~
	\label{constraint New1}
	& \text{tr}\left(\mathbf{\bar{H}}_\text{E} \left(\mathbf{\Theta}_\text{E}\right) \mathbf{\tilde{\Theta}}_\text{B}\right) + \lambda \left(\mathbf{\bar{h}}_\text{E}\left(\mathbf{\Theta}_\text{E}\right) + 1\right) = 1
	\\
	\label{constraint New2}
	& \text{tr}\left(\mathbf{E}_m \mathbf{\tilde{\Theta}}_\text{B}\right)=\lambda,~\forall m \in \mathcal{N}_\text{B},\,\mathbf{\tilde{\Theta}}_\text{B} \succcurlyeq 0,~\lambda \geq 0.
\end{align}
\end{subequations}
Problem \eqref{problem theta 4} is a semidefinite program (SDP) which is convex, and can be optimally solved by using the convex solvers, e.g., the interior point methods \cite{6728676}. After finding the optimal $\mathbf{\tilde{\Theta}}_\text{B}$, we obtain $\boldsymbol{\bar{\Theta}}_\text{B} = \frac{1}{\lambda} \mathbf{\tilde{\Theta}}_\text{B}$. Then, due to constraint $\text{rank}(\boldsymbol{\bar{\Theta}}_\text{B})=1$, we apply the standard Gaussian randomization method and obtain $\boldsymbol{\bar{\theta}}_\text{B}$. Finally, $\boldsymbol{\theta}_\text{B}$ can be obtained by $\boldsymbol{\bar{\theta}}_\text{B} = \left[ \boldsymbol{\theta}^T_\text{B},1 \right]^T$.

The output of the proposed algorithm for solving \eqref{problem theta 4} may be infeasible, due to the relaxation of constraints. In this line, we apply the quantization method based on the Euclidean distance \cite{9446526,9343768}. In this method, we apply the quantization method to each random vector generated via the Gaussian randomization algorithm. The quantization method is described as follows:
Let us denote a generated vector by the Gaussian randomization algorithm as $\boldsymbol{\theta}^\text{(0)}_\text{B}=[\theta^{\text{B},(0)}_m],~\forall m \in \mathcal{N}_\text{B}$. The feasible $\boldsymbol{\theta}^\text{(1)}_\text{B}=[\theta^{\text{B},(1)}_m],~\forall m \in \mathcal{N}_\text{B}$ can be obtained as
$\theta^{\text{B},(1)}_m=\frac{2k^*_n\pi}{L_\text{B}},~\forall m \in \mathcal{N}_\text{B}$, where $k^*_m = \argmin\limits_{k=0,\dots,(L_\text{B}-1)} \left\lvert \theta^{\text{B},(0)}_m - e^{j \frac{2k\pi}{L_\text{B}}} \right\rvert$. In this quantized Gaussian randomization method, we choose the vector $\boldsymbol{\theta}^\text{(1)}_\text{B}$ which leads to the maximum secrecy rate formulated in \eqref{secrecy rate}.

\subsubsection{Finding $\mathbf{w}$}\label{subsec opt w}
For any given $(\mathbf{\Theta}_\text{B},\mathbf{\Theta}_\text{E})$, problem \eqref{main problem}  can be equivalently transformed to the following form
\begin{align}\label{problem 1}
	\max\limits_{\mathbf{w}}~
	\frac{\mathbf{w}^H \mathbf{\tilde{H}}_\text{B}\left(\mathbf{\Theta}_\text{B},\mathbf{\Theta}_\text{E}\right) \mathbf{w} + 1}
	{\mathbf{w}^H \mathbf{\tilde{H}}_\text{E}\left(\mathbf{\Theta}_\text{B},\mathbf{\Theta}_\text{E}\right) \mathbf{w} + 1}~~~~~~
	\text{s.t.}~~
	\mathbf{w}^H \mathbf{w} \leq P,
\end{align}
where $\mathbf{\tilde{H}}_\text{B} \left(\mathbf{\Theta}_\text{B},\mathbf{\Theta}_\text{E}\right) = \frac{1}{\sigma^2_B} \mathbf{h}^H_\text{B}\left(\mathbf{\Theta}_\text{B},\mathbf{\Theta}_\text{E}\right) \mathbf{h}_\text{B}\left(\mathbf{\Theta}_\text{B},\mathbf{\Theta}_\text{E}\right)$, $\mathbf{\tilde{H}}_\text{E} \left(\mathbf{\Theta}_\text{B},\mathbf{\Theta}_\text{E}\right) = \frac{1}{\sigma^2_E} \mathbf{h}^H_\text{E}\left(\mathbf{\Theta}_\text{B},\mathbf{\Theta}_\text{E}\right) \mathbf{h}_\text{E}\left(\mathbf{\Theta}_\text{B},\mathbf{\Theta}_\text{E}\right)$, in which the superscript $H$ denotes the conjugate transpose operation. The optimal beamforming $\mathbf{w}_\text{opt}$ for problem \eqref{problem 1} can be obtained in closed form as follows:
\begin{align}\label{EQ_opt_w}
	\mathbf{w}_\text{opt} = \sqrt{P} \boldsymbol{u}_\text{max},
\end{align}
where $\boldsymbol{u}_\text{max}$ is the normalized eigenvector corresponding to the largest eigenvalue of $(\mathbf{\tilde{H}}_\text{E}\left(\mathbf{\Theta}_\text{B},\mathbf{\Theta}_\text{E}\right) + \frac{1}{P} \mathbf{I}_\text{M})^{-1} (\mathbf{\tilde{H}}_\text{B}\left(\mathbf{\Theta}_\text{B},\mathbf{\Theta}_\text{E}\right) + \frac{1}{P} \mathbf{I}_\text{M})$, in which $\mathbf{I}_\text{M}$ denotes an $M \times M$ identity matrix \cite{5485016}.

\subsubsection{Finding $\mathbf{\Theta}_\text{E}$}\label{subsec find thetaE}
For any given $\left(\mathbf{w},\mathbf{\Theta}_\text{B}\right)$, problem \eqref{main problem} can be rewritten as
\begin{align}\label{problem phi 1}
	\min\limits_{\mathbf{\Theta}_\text{E}}~
	\frac{\frac{1}{\sigma^2_B} \lvert \mathbf{h}_\text{B}\left(\mathbf{\Theta}_\text{B},\mathbf{\Theta}_\text{E}\right) \mathbf{w} \rvert^2 + 1}
	{\frac{1}{\sigma^2_E} \lvert \mathbf{h}_\text{E}\left(\mathbf{\Theta}_\text{B},\mathbf{\Theta}_\text{E}\right) \mathbf{w} \rvert^2 + 1}~~~~~~
	\text{s.t.}~\eqref{constraint phi}.
\end{align}
The resulting problem \eqref{problem phi 1} has a similar structure to \eqref{problem 1a}. Therefore, the proposed algorithm for solving \eqref{problem 1a} can be easily modified to be applied to \eqref{problem phi 1}. To avoid duplication, the details of the modified algorithm for solving problem \eqref{problem phi 1} is not presented in the paper.

\subsubsection{Initialization Method}\label{subsec initial}
Here, we initialize $\mathbf{\Theta}_\text{B}$, $\mathbf{w}$, and $\mathbf{\Theta}_\text{E}$, respectively, by using a low-complexity, yet suboptimal, method. For initializing $\mathbf{\Theta}_\text{B}$, we assume that $\mathbf{w}$ and $\mathbf{\Theta}_\text{E}$ are unknown, thus $\mathbf{\Theta}_\text{B}$ is designed to maximize the effective channel of Bob, denoted by\footnote{In the simplified Bob's effective channel formulation, the term $\mathbf{h}_\text{$\text{I}_\text{E}$,B} \mathbf{\Theta}_\text{E} \mathbf{h}_\text{A,$\text{I}_\text{E}$}$ in \eqref{signal Bob} is ignored, since in this step, $\mathbf{\Theta}_\text{E}$ is unknown.} $\mathbf{h}_\text{B}\left(\mathbf{\Theta}_\text{B}\right) = \left( \mathbf{h}_\text{A,B} + \mathbf{h}_\text{$\text{I}_\text{B}$,B} \mathbf{\Theta}_\text{B} \mathbf{h}_\text{A,$\text{I}_\text{B}$} \right)$. The resulting problem is given by
\begin{align}\label{problem initial thetaB}
	\max\limits_{\mathbf{\Theta}_\text{B}}~
	\norm{\mathbf{h}_\text{A,B} + \mathbf{h}_\text{$\text{I}_\text{B}$,B} \mathbf{\Theta}_\text{B} \mathbf{h}_\text{A,$\text{I}_\text{B}$}}^2~~~~~~
	\text{s.t.}~\eqref{constraint theta}.
\end{align}
To solve \eqref{problem initial thetaB}, we first relax the integer constraint \eqref{constraint theta} by replacing it with the unit-norm constraint $|\theta^\text{B}_m| = 1,~\forall m \in \mathcal{N}_\text{B}$. The resulting problem has a similar structure to problem (13) in \cite{8811733}. Therefore, the proposed method in \cite{8811733} is utilized to solve \eqref{problem initial thetaB} and find efficient $\mathbf{\Theta}_\text{B}$. To meet constraint \eqref{constraint theta}, we employ the quantized Gaussian randomization method described in Subsection \ref{subsec find thetaB}.

Similar to $\mathbf{\Theta}_\text{B}$, we find $\mathbf{\Theta}_\text{E}$ such that the norm of the effective channel of Eve, denoted by  $\mathbf{h}_\text{E}\left(\mathbf{\Theta}_\text{E}\right) = \left( \mathbf{h}_\text{A,E} + \mathbf{h}_\text{$\text{I}_\text{E}$,E} \mathbf{\Theta}_\text{E} \mathbf{h}_\text{A,$\text{I}_\text{E}$} \right)$, is maximized. The resulting problem is 
\begin{align}\label{problem initial thetaE}
	\max\limits_{\mathbf{\Theta}_\text{E}}~
	\norm{\mathbf{h}_\text{A,E} + \mathbf{h}_\text{$\text{I}_\text{E}$,E} \mathbf{\Theta}_\text{E} \mathbf{h}_\text{A,$\text{I}_\text{E}$}}^2~~~~~~
	\text{s.t.}~\eqref{constraint phi},
\end{align}
which has a similar structure to problem \eqref{problem initial thetaB}. Hence, the utilized method for solving \eqref{problem initial thetaB} can be modified for solving \eqref{problem initial thetaE}. 

After initializing $\mathbf{\Theta}_\text{B}$ and $\mathbf{\Theta}_\text{E}$, we initialize $\mathbf{w}$ by using the optimal closed-form expression \eqref{EQ_opt_w}. Finally, to consider the worst-case secrecy rate, we update $\mathbf{\Theta}_\text{E}$ for the given $\mathbf{\Theta}_\text{B}$ and $\mathbf{w}$ by using our proposed method in Subsection \ref{subsec find thetaE}. The pseudo code of the proposed AO-based method for solving \eqref{main problem} is described in Alg. \ref{Alg_AO}.
\begin{algorithm}[tp]
	\caption{The alternating optimization method.}
	\label{Alg_AO}
	\begin{algorithmic}[1]
		\STATE Initialize $\mathbf{\Theta}_\text{B}$, $\mathbf{w}$, and $\mathbf{\Theta}_\text{E}$ by solving \eqref{problem phi 1}. Set maximum iterations $T$, and tolerance $\epsilon$.
		\FOR {$l=1:L$}
		\STATE Find $\boldsymbol{\bar{\theta}}_\text{B}$ by solving problem \eqref{problem theta 4}. Then, update $\boldsymbol{\theta}_\text{B}$ by using $\boldsymbol{\bar{\theta}}_\text{B} = \left[ \boldsymbol{\theta}^T_\text{B},1 \right]^T$.
		\\
		\STATE Update the beamforming vector $\mathbf{w}$ according to \eqref{EQ_opt_w}.
		\\
		\STATE Update $\boldsymbol{\theta}_\text{E}$ according to Subsection \ref{subsec find thetaE}.
		\\
		\STATE\textbf{if}~~ $|C^{(l)}_\text{s} - C^{(l-1)}_\text{s}| \leq \epsilon$~~\textbf{then}
		\\
		\STATE~~~\textbf{break}.
		\\
		\STATE\textbf{end if}
		\ENDFOR
	\end{algorithmic}
\end{algorithm}

\subsection{Gradient Descent Ascent}
To invoke GDA, which is used to solve min-max problems, we first transform the max-min problem \eqref{main problem} into a min-max form as follows:
\begin{subequations}
	\begin{align}\label{obf main2}
		\min\limits_{\mathbf{w},\mathbf{\Theta}_\text{B}}~\hspace{.0 cm} 
		\max\limits_{\mathbf{\Theta}_\text{E}}  &\quad\log_2 \left( 1 + \frac{ \lvert \mathbf{H}_\text{E}\left(\mathbf{\Theta}_\text{B},\mathbf{\Theta}_\text{E}\right) \mathbf{w} \rvert^2}{\sigma^2_E} \right)-\notag\\
		&\hspace{1.5cm}\log_2 \left( 1 + \frac{ \lvert \mathbf{H}_\text{B}\left(\mathbf{\Theta}_\text{B},\mathbf{\Theta}_\text{E}\right) \mathbf{w} \rvert^2}{\sigma^2_B} \right) \\
		\text{s.t.}&\quad
		\eqref{constraint power}\text{-}\eqref{constraint phi}.\nonumber
	\end{align}
\end{subequations}
Following the same steps to formulate \eqref{problem theta 4}, we define $\lambda_2 = \left(\text{tr}\left(\mathbf{\bar{H}}_\text{B2} \left(\mathbf{\Theta}_\text{B}\right) \mathbf{\tilde{\Theta}}_\text{E}\right) +  \mathbf{\bar{h}}_\text{B2}\left(\mathbf{\Theta}_\text{B}\right) + 1\right)^{-1}$ and $\mathbf{\tilde{\Theta}}_\text{E} = \lambda_2 \boldsymbol{\bar{\Theta}}_\text{E}$ and then
formulate a linearized inner optimization problem with respect to $\mathbf{\tilde{\Theta}}_\text{E}$ and $\lambda_2$ as follows:
\begin{subequations}\label{problem theta 42}
	\begin{align}\label{obf problem theta 42}
		\hspace{-0.3cm}\max\limits_{\mathbf{\tilde{\Theta}}_\text{E} ,~\lambda_2 }~
		& f(\mathbf{\Theta_B},\mathbf{\tilde{\Theta}_E},\mathbf{w},\lambda_2)  \\
		\text{s.t.}~~
		\label{constraint New12}
		& \text{tr}\left(\mathbf{\bar{H}}_\text{B2} \left(\mathbf{\Theta}_\text{B}\right) \mathbf{\tilde{\Theta}}_\text{E}\right) + \lambda_2 \left(\mathbf{\bar{h}}_\text{B2}\left(\mathbf{\Theta}_\text{B}\right) + 1\right) = 1
		\\
		\label{constraint New22}
		& \text{tr}\left(\tilde{\mathbf{E}}_n \mathbf{\tilde{\Theta}}_\text{E}\right)=\lambda_2,~\forall n \in \mathcal{N}_\text{E}\\
		& \mathbf{\tilde{\Theta}}_\text{E} \succcurlyeq 0,~\lambda_2 \geq 0,
	\end{align}
\end{subequations}
where
\begin{align}
	&f(\!\mathbf{\Theta_B},\mathbf{\tilde{\Theta}_\text{E}},\mathbf{w},\lambda_2\!)\!=\!
	\text{tr}\!\left(\!\mathbf{\bar{H}}_\text{E2}\! \left(\!\mathbf{\Theta}_\text{B}\!\right) \! \mathbf{\tilde{\Theta}}_\text{E}\!\right) \!+\! \lambda_2 \!\left(\mathbf{\bar{h}}_\text{E2} \left(\!\mathbf{\Theta}_\text{B}\!\right) \!+\! 1\right),\\
	&\mathbf{\tilde{\Theta}}_\text{E} = \lambda_2 \boldsymbol{\bar{\theta}}_\text{E}\boldsymbol{\bar{\theta}}_\text{E}^H,\,\, \boldsymbol{\bar{\theta}}_\text{E} = \left[ \boldsymbol{\theta}^T_\text{E} , 1 \right]^T,\,\,\mathbf{Q}= \mathbf{w} \mathbf{w}^H,\\
	&\mathbf{\bar{h}}_\text{E2} = \sigma^{-2}_E\left( \mathbf{h}_\text{A,E} + \mathbf{h}_\text{$\text{I}_\text{B}$,E} \mathbf{\Theta}_\text{B} \mathbf{h}_\text{A,$\text{I}_\text{B}$} \right) \mathbf{Q} \left( \mathbf{h}_\text{A,E} + \mathbf{h}_\text{$\text{I}_\text{B}$,E} \mathbf{\Theta}_\text{B} \mathbf{h}_\text{A,$\text{I}_\text{B}$} \right)^H, \label{EQ_he2}\\
	&\mathbf{\bar{h}}_\text{B2} = \sigma^{-2}_B\left( \mathbf{h}_\text{A,B} + \mathbf{h}_\text{$\text{I}_\text{B}$,B} \mathbf{\Theta}_\text{B} \mathbf{h}_\text{A,$\text{I}_\text{B}$} \right) \mathbf{Q} \left( \mathbf{h}_\text{A,B} + \mathbf{h}_\text{$\text{I}_\text{B}$,B} \mathbf{\Theta}_\text{B} \mathbf{h}_\text{A,$\text{I}_\text{B}$} \right)^H,
\end{align}
and $\mathbf{\bar{H}}_\text{B2}  \left(\mathbf{\Theta}_\text{B}\right)$ and $\mathbf{\bar{H}}_\text{E2}  \left(\mathbf{\Theta}_\text{B}\right)$ are defined as \eqref{Hbar B2} and \eqref{Hbar E2}, respectively. 
\begin{figure*}[!t]
	\begin{align}
		&\mathbf{\bar{H}}_\text{B2}  \left(\mathbf{\Theta}_\text{B}\right)= \frac{1}{\sigma^2_B} \begin{bmatrix} \text{diag} \left( \mathbf{h}^*_\text{$\text{I}_\text{E}$,B} \right) \mathbf{h}^*_\text{A,$\text{I}_\text{E}$} \mathbf{w}^* \mathbf{w}^T \mathbf{h}^T_\text{A,$\text{I}_\text{E}$} \text{diag} \left( \mathbf{h}_\text{$\text{I}_\text{E}$,B} \right) & \text{diag} \left( \mathbf{h}^*_\text{$\text{I}_\text{E}$,B} \right) \mathbf{h}^*_\text{A,$\text{I}_\text{E}$} \mathbf{w}^* \mathbf{w}^T \left( \mathbf{h}^T_\text{A,B} + \mathbf{h}^T_\text{A,$\text{I}_\text{E}$} \mathbf{\Theta}_\text{B} \mathbf{h}^T_\text{$\text{I}_\text{E}$,B} \right) \\
			\left( \mathbf{h}^*_\text{A,B} + \mathbf{h}^*_\text{$\text{I}_\text{E}$,B} \mathbf{\Theta}_\text{B}^* \mathbf{h}^*_\text{A,$\text{I}_\text{E}$} \right) \mathbf{w}^* \mathbf{w}^T \mathbf{h}_\text{$A,\text{I}_\text{E}$}^T\text{diag} \left( \mathbf{h}_\text{$\text{I}_\text{E}$,B} \right) & 0 \end{bmatrix}, \label{Hbar B2}
		\\
		&\mathbf{\bar{H}}_\text{E2}  \left(\mathbf{\Theta}_\text{B}\right)= \frac{1}{\sigma^2_E} \begin{bmatrix} \text{diag} \left( \mathbf{h}^*_\text{$\text{I}_\text{E}$,E} \right) \mathbf{h}^*_\text{A,$\text{I}_\text{E}$} \mathbf{w}^* \mathbf{w}^T \mathbf{h}^T_\text{A,$\text{I}_\text{E}$} \text{diag} \left( \mathbf{h}_\text{$\text{I}_\text{E}$,E} \right) 
			& \text{diag} \left( \mathbf{h}^*_\text{$\text{I}_\text{E}$,E} \right) \mathbf{h}^*_\text{A,$\text{I}_\text{E}$} \mathbf{w}^* \mathbf{w}^T \left( \mathbf{h}^T_\text{A,E} + \mathbf{h}^T_\text{A,$\text{I}_\text{B}$} \mathbf{\Theta}_\text{B} \mathbf{h}^T_\text{$\text{I}_\text{B}$,E} \right) \\
			\left( \mathbf{h}^*_\text{A,E} + \mathbf{h}^*_\text{$\text{I}_\text{B}$,E} \mathbf{\Theta}_\text{B}^* \mathbf{h}^*_\text{A,$\text{I}_\text{B}$} \right) \mathbf{w}^* \mathbf{w}^T \mathbf{h}_\text{$A,\text{I}_\text{E}$}^T\text{diag} \left( \mathbf{h}_\text{$\text{I}_\text{E}$,E} \right) & 0, \end{bmatrix}, \label{Hbar E2}
	\end{align}
\end{figure*}
The parameter $\tilde{\mathbf{E}}_n$ is an $(N_\text{E}+1) \times (N_\text{E}+1)$ diagonal matrix, defined similar to \eqref{rate2brac}.
The GDA scheme includes the following three main steps:

\textbf{Step 1}: Given $\mathbf{\Theta}_\text{B}$, $\mathbf{w}$, $\tilde{\mathbf{{\Theta}}}_\text{E}$, and $\lambda_2$ from the $r$-th round, we can update $\mathbf{{\Theta}}_\text{B}$ in the $(r+1)$-th round by GDA as follows:
\begin{align}\label{EQ_B_iter}
	&\mathbf{\Theta}^{(r+1)}_\text{B}=\nonumber\\
	&\mathcal{P}^{(r+1)}_{\mathcal{T}_\text{B}} \left( \mathbf{\Theta}^{(r)}_\text{B}
	-\alpha\nabla_{\mathbf{ {\Theta}_\text{B}}}f\left(\mathbf{\tilde{\Theta}_\text{E}}^{(r)},{\boldsymbol{\Theta}_\text{B}}^{(r)},\mathbf{w}^{(r)},\lambda_2^{(r)}\right)\right),
\end{align}
where (for more details, please see Appendix \ref{APP_gradient_ThB})
\begin{multline}\label{EQ_grad_TB}
	\nabla_{\mathbf{\Theta}_\text{B}} f = {\lambda_2}\bigg(\mathbf{h}_\text{A,$\text{I}_\text{B}$}^*\mathbf{A}^H \boldsymbol{\theta_\text{E}} \mathbf{h}_\text{$\text{I}_\text{B}$,E}^*+ \\
	\frac{1}{\sigma_\text{E}^2}\mathbf{h}_\text{$\text{I}_\text{B}$,E}^H\left( \mathbf{h}_\text{A,\text{E}} +\mathbf{h}_\text{$\text{I}_\text{B}$,E} \mathbf{\Theta}_\text{B} \mathbf{h}_\text{A,$\text{I}_\text{B}$} \right) \mathbf{Q}^T\mathbf{h}_\text{A,$\text{I}_\text{B}$}^H\bigg),
\end{multline}
$\mathbf{A}=\text{diag} \left( \mathbf{h}^*_\text{$\text{I}_\text{E}$,E} \right) \mathbf{h}^*_\text{A,$\text{I}_\text{E}$} \mathbf{w}^* \mathbf{w}^T$,
\begin{align}\label{EQ_Proj_TB}
	&\mathcal{P}_{\mathcal{T}_\text{B}}^{(r+1)}\left(\boldsymbol{ {\Theta}_\text{B0}}\right)=\argmin_{\boldsymbol{ {\Theta}_\text{B}}\in\mathcal{T}_\text{B}^{(r+1)} } ||\boldsymbol{ {\Theta}_\text{B}}-\mathbf{ {\Theta}_\text{B0}}||_F^2, 
\end{align}
in which
$\mathcal{T}_\text{B}^{(r+1)}=\bigg\{\boldsymbol{ {\Theta}_\text{B}}:\,\mbox{tr}\left(\hat{\mathbf{E}}_n \boldsymbol{{\Theta}}_\text{B}\right)=\lambda_2^{(r)},~\forall n \in \mathcal{N}_\text{B},\,\boldsymbol{{\Theta}}_\text{B}\succeq \mathbf{0}, \mbox{tr}\left(\mathbf{\bar{H}}_\text{B2} \left(\boldsymbol{\Theta_\text{B}}\right) \boldsymbol{\tilde{\Theta}}_\text{E}^{(r)}\right) \!+\! \lambda_2^{(r)} \left(\mathbf{\bar{h}}_\text{B2}^{}\left(\boldsymbol{\Theta_\text{B}}^{}\right) \!+\! 1\right) = 1\!\bigg\}$, and
$\hat{\mathbf{E}}_n$ is an $N_\text{B} \times N_\text{B}$ diagonal matrix, defined similar to \eqref{rate2brac} except the last diagonal term. The optimization problem \eqref{EQ_Proj_TB} has a quadratic objective function with linear constraints, thus it can be solved by using the standard convex solvers.

\textbf{Step 2}: After updating $\mathbf{{\Theta}}_\text{B}$, we update $\mathbf{w}$ by using \eqref{EQ_opt_w}.

\textbf{Step 3}: Given $\mathbf{{\Theta}}_\text{B}$ and $\mathbf{w}$ from the $(r+1)$-th round and also $\mathbf{\tilde{{\Theta}}}_\text{E}$ and $\lambda_2$ from the $r$-th round, we can update  $\mathbf{\tilde{\Theta}_\text{E}}$ and $\lambda_2$ in the $(r+1)$-th round as follows:
\begin{align}\label{EQ_TE_iter2}
	&(\!\mathbf{\tilde{\Theta}_\text{E}}^{\!(r\!+\!1)}\!\!,\!{\lambda_2}^{\!(r\!+\!1)}\!)\!=\!\mathcal{P}_{\mathcal{T}_\text{E}}^{(r+1)}\!\!\left(\!\mathbf{\tilde{\Theta}_\text{E}}^{(r)}\!\!\!+\!\!\alpha\!\nabla_{\!\!\mathbf{\tilde{\Theta}_\text{E}}}\!f\!\!\left(\!\mathbf{\Theta_\text{B}}^{(r+1)}\!\!,\!\mathbf{\tilde{\Theta}_\text{E}}^{(r)}\!,\!\mathbf{w}^{(r)}\!,\!\lambda_2^{(r)}\!\right)\!,\!\right.\notag\\
	&\hspace{1.3cm}\left.{\lambda_2}^{(r)}+\alpha\nabla_{ \lambda_2}f\left(\mathbf{\tilde{\Theta}_\text{E}}^{(r)},\boldsymbol{ {\Theta}_\text{B}}^{(r+1)},\mathbf{w}^{(r)},\lambda_2^{(r)}\right)\right)
	\notag\\
	&=\mathcal{P}_{\mathcal{T}_E}^{(r+1)}\left(\mathbf{\tilde{\Theta}_\text{E}}^{(r)},{\lambda_2}^{(r)}+\alpha\left(\mathbf{\bar{h}}_\text{E2} \left(\mathbf{\Theta}_\text{B}^{(r+1)}\right) + 1\right)\right),
\end{align}
where \eqref{EQ_TE_iter2} is from \cite[Table 4.3]{hjorungnes_2011} and 
$\nabla_{\mathbf{\tilde{\Theta}_\text{E}}}f=\frac{df}{d\tilde{\Theta}_\text{E}^*}$ with the fact that $\frac{d\mbox{tr}(\mathbf{A}\mathbf{\tilde{\Theta}_\text{E}})}{d\mathbf{\tilde{\Theta}_\text{E}}^*}=0$, the projection in \eqref{EQ_TE_iter2} is defined as:
\begin{subequations}
\begin{align}
&\mathcal{P}_{\mathcal{T}_E}^{(r+1)} \left(\mathbf{\tilde{\Theta}_\text{E0}},\,\lambda_0\right)=\!\!\!\argmin_{(\mathbf{\tilde{\Theta}_\text{E},\,\lambda_2)}\in\mathcal{T}_\text{E}^{(r+1)}}  \!\!\!||\boldsymbol{\tilde{\Theta}_\text{E}}\!-\!\mathbf{\tilde{\Theta}_\text{E0}}||_F^2+|\lambda_2\!-\!\lambda_0|^2, \notag
\end{align}
where the set $\mathcal{T}_E^{(r+1)}$ is defined as follows:
\begin{align}
&\hspace{0cm}\mathcal{T}_E^{(r+1)}=\bigg\{\left(\boldsymbol{\tilde{\Theta}_\text{E}},\lambda_2\right):\quad\notag\\
&\hspace{0.0cm}\mbox{tr}\left(\mathbf{\bar{H}}_\text{B2} \left(\boldsymbol{\tilde{\Theta}_\text{B}}^{(r+1)}\right) \boldsymbol{\tilde{\Theta}}_\text{E}\right) + \lambda_2 \left(\mathbf{\bar{h}}_\text{B2}\left(\boldsymbol{\tilde{\Theta}_\text{B}}^{(r+1)}\right) + 1\right) = 1,\label{EQ_Proj_E_constr1}  \\
&\hspace{0cm} \mbox{tr}\left(\tilde{\mathbf{E}}_n \boldsymbol{\tilde{\Theta}}_\text{E}\right)=\lambda_2,~\forall n \in \mathcal{N}_\text{E},\,\boldsymbol{\tilde{\Theta}}_\text{E}\succeq \mathbf{0},\,\lambda_2\geq 0\quad\bigg\},\label{EQ_constr_unit_mag_tE}
\end{align}
\end{subequations}
where $||.||_F^2$ is the Frobenius norm, \eqref{EQ_Proj_E_constr1} is from the CCT and \eqref{EQ_constr_unit_mag_tE} is from the unit magnitude constraint. Then we iteratively run Steps 1-3 until the algorithm converges. The pseudo code of our proposed GDA-based method is presented in Alg. \ref{Alg_SO}.
\begin{algorithm}[tp]
	\caption{The gradient descent ascent algorithm.}
	\label{Alg_SO}
	\begin{algorithmic}[1]
		\STATE Initialize $\mathbf{\Theta}_\text{B}$, $\mathbf{w}$, and $\mathbf{\Theta}_\text{E}$ according to Subsection \ref{subsec initial}. Set maximum iterations $T$, and tolerance $\epsilon$.
		\FOR {$l=1:L$}
		\STATE Update $\boldsymbol{\theta}_\text{B}$ by solving problem \eqref{EQ_B_iter}.
		\\
		\STATE Update the beamforming vector $\mathbf{w}$ according to \eqref{EQ_opt_w}.
		\\
		\STATE Update $\boldsymbol{\theta}_\text{E}$ by solving problem \eqref{EQ_TE_iter2}
		\\
		\STATE\textbf{if}~~ $|C^{(l)}_\text{s} - C^{(l-1)}_\text{s}| \leq \epsilon$~~\textbf{then}
		\\
		\STATE~~~\textbf{break}.
		\\
		\STATE\textbf{end if}
		\ENDFOR
	\end{algorithmic}
\end{algorithm}

\subsection{Game Theoretical Method}
We review the games in strategic form where players choose their strategy once and simultaneously with all other players without knowing the others' actions. The game $\Gamma$ can be described by the tuple $\Gamma = (\mathcal{N}, \mathcal{S}, \bf{u})$ with $\mathcal{N}=\{\text{Bob}, \text{Eve}\}$ denoting the set of players, $\mathcal{S}$ the joint strategy space, and $\bf{u}$ the utility function. Since the players, Bob's IRS and Eve's IRS, have conflicting interests, i.e., Bob's and Eve's IRS aim to maximize and minimize the secrecy rate, respectively, this game can be modeled as a two-player zero-sum game. We represent the players' utilities by a matrix, which is defined as $\mathbf{A}=[a_{ij}],~\forall i=1,\dots,L_\text{B}^{N_{\text{B}}},~j=1,\dots,L_\text{E}^{N_{\text{E}}}$, where
$a_{ij}$ and $-a_{ij}$ denote the utilities of Bob's and Eve's IRS, respectively. The min-max value is equal to the max-min value in any finite two-player zero-sum game which corresponds to a Nash equilibrium (NE). The NE is a strategy profile in which all players choose the best response of the other players' strategies. In the mixed strategies, the player chooses her actions randomly and independently of the other players' choices according to a probability distribution. Note that there exists a NE equilibrium in the mixed strategy in any game with finite set of players with a finite set of actions. Therefore, we have at least one NE in our strategic game. We can then compute the mixed NE strategy $\mathbf{x}=[x_i],\forall i=1, \dots, L_B^{N_B}$, for Bob's IRS by solving a linear program \cite{lemke}.
Similarly, Eve's IRS could also randomize her actions by the same procedure and obtain her mixed NE strategy $\mathbf{y}=[y_j],\forall j=1, \dots, L_E^{N_E}$, such that none of the players would gain a higher payoff by deviating unilaterally from their NE strategy.

\section{Numerical results}
\begin{figure*}
	\centering
	\subfigure[Network topology and placement of elements.]{
		\includegraphics[scale=0.47]{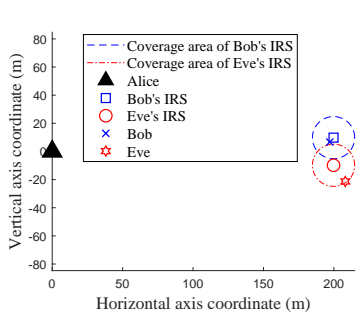}
		\label{Fig_topology}
	}\hfil
	\subfigure[Achievable rate vs. iteration index: The continuous domain for phase shifting elements.]{
		\includegraphics[scale=0.38]{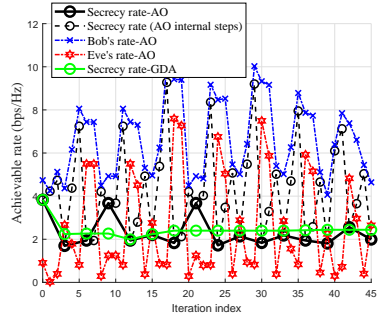}
		\label{Fig_SR_FirstBob}
	}\hfil
	\subfigure[Achievable rate vs. iteration index: The discrete domain for phase shifting elements.]{
		\includegraphics[scale=0.38]{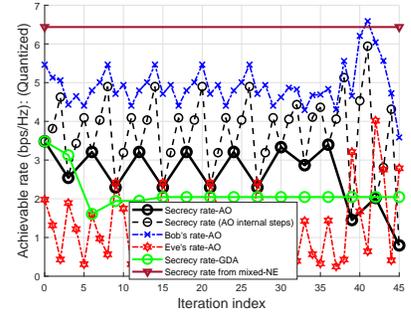}
		\label{Fig_SR_FirstBob_quantized}
	}\hfil
	\caption
	{The secrecy/receivers data rate over AO and GDA iterations.}
	\label{SimulationFigures}
\end{figure*}
We evaluate the convergence and performance of AO and GDA. We set $M=3$, $N_\text{B}=N_\text{E}=4$, $P=46$ dBm, $5$ MHz wireless band, and AWGN power density $-174$ dBm/Hz. The path loss exponents for the direct and reflected channels are set to $4$ and $2$, respectively. We apply the MIMO channel correlation model in \cite{1033686} for the channels between Alice to each IRS. In the following, we consider the case that these MIMO channels are full-rank.
For Gaussian randomization, we generate $10,000$ random vectors according to the optimized covariance matrix.

Fig. \ref{Fig_SR_FirstBob} shows the convergence behavior of AO and GDA for the continuous domain of phase shifting strategies. Within each AO iteration, we observe that after optimizing $\mathbf{w}$ and $\Theta_\text{B}$, the secrecy rate is non-decreasing. On the other hand, the secrecy rate is non-increasing after optimizing $\Theta_\text{E}$ at each AO iteration, verifying the feasibility of our adopted SDR techniques.
Although AO may achieve better performance than GDA in some iterations, it does not guarantee convergence in general. This is due to the fact that for any given $\mathbf{w}$ and $\Theta_\text{B}$, optimizing $\Theta_\text{E}$ effectively changes the phase of Eve's equivalent channel $\mathbf{h}_\text{E}\left(\mathbf{\Theta}_\text{B},\mathbf{\Theta}_\text{E}\right)$, specifically when the norm of Eve's IRS channel dominates other channels, i.e., 
$\norm {\mathbf{h}_\text{$\text{I}_\text{E}$,E} \mathbf{\Theta}_\text{E} \mathbf{h}_\text{A,$\text{I}_\text{E}$}} \gg \norm{ \mathbf{h}_\text{A,E}}$, and $\norm{\mathbf{h}_\text{$\text{I}_\text{E}$,E} \mathbf{\Theta}_\text{E} \mathbf{h}_\text{A,$\text{I}_\text{E}$}} \gg \norm{\mathbf{h}_\text{$\text{I}_\text{B}$,E} \mathbf{\Theta}_\text{B} \mathbf{h}_\text{A,$\text{I}_\text{B}$}}$. When Bob is close to Eve's IRS, $\mathbf{\Theta}_\text{E}$ can effectively change the phase of Bob's equivalent channel $\mathbf{h}_\text{B}\left(\mathbf{\Theta}_\text{B},\mathbf{\Theta}_\text{E}\right)$, 
such that it seriously degrades secrecy rate, due to the fixed $\mathbf{w}$. Hence, when $\norm{\mathbf{h}_\text{$\text{I}_\text{E}$,B} \mathbf{\Theta}_\text{E} \mathbf{h}_\text{A,$\text{I}_\text{E}$}} \gg \norm{\mathbf{h}_\text{A,B}}$, and $\norm{\mathbf{h}_\text{$\text{I}_\text{E}$,B} \mathbf{\Theta}_\text{E} \mathbf{h}_\text{A,$\text{I}_\text{E}$}} \gg \norm{\mathbf{h}_\text{$\text{I}_\text{B}$,B} \mathbf{\Theta}_\text{B} \mathbf{h}_\text{A,$\text{I}_\text{B}$}}$, we observe that Eve's IRS may act as 
a jammer to Bob by changing the phase of $\mathbf{h}_\text{B}\left(\mathbf{\Theta}_\text{B},\mathbf{\Theta}_\text{E}\right)$ rather than enhancing $\mathbf{h}_\text{E}\left(\mathbf{\Theta}_\text{B},\mathbf{\Theta}_\text{E}\right)$. A similar effect can be observed at Bob's IRS. The convergence behavior of GDA is smoother than AO such that after few iterations, it converges to a stationary point.
Fig. \ref{Fig_SR_FirstBob} shows the convergence behavior of AO and GDA for $L_\text{B}=L_\text{E}=5$. Discrete domains for $\mathbf{\Theta}_\text{B}$ and $\mathbf{\Theta}_\text{E}$ usually improves the convergence of AO. This effect is because $\mathbf{\Theta}_\text{E}$ cannot be effectively chosen under a discrete domain. Although the same argument holds for $\mathbf{\Theta}_\text{B}$, the beamforming at Alice with a continuous domain can effectively compensate the inflexibility of choosing $\mathbf{\Theta}_\text{B}$. From numerical results, we observed that AO with discrete domains for $\mathbf{\Theta}_\text{B}$ and $\mathbf{\Theta}_\text{E}$ has smoother convergence than that of the continuous one, although it does not always hold.
In the game-theoretic approach, the game has a single mixed NE. At that mixed NE, it is interesting to note that only five phase shift combinations share the whole probability mass of the optimal strategy for Bob and Eve. Following that strategy yields a mixed secrecy capacity of $6.436$ bps/Hz, which is higher than those of AO and GDA. In the game-theoretic approach, the priority of updating $\mathbf{\Theta}_\text{E}$ after optimizing $\mathbf{w}$ is not considered, so the optimal beamforming can effectively deteriorate Eve's rate. 

\section{Conclusion}


We consider a wiretap channel where both the legitimate receiver and the eavesdropper use their own IRSs. We formulate a max-min secrecy rate problem, to design the transmit beamforming and phase shifting strategies of the IRSs by AO, GDA, and mixed NE. Simulation results show that AO does not guarantee convergence for continuous phase shifting, although it may achieve better performance than GDA in some iterations. GDA usually converges to a stationary point. Discrete phase shifting improves the convergence behavior of AO and GDA. The mixed NE strategy (with no priority among the players) achieves higher secrecy rate compared to AO and GDA.

\appendices

\section{Derivation of $\nabla_{\mathbf{\Theta}_\text{B}}f$ in \eqref{EQ_B_iter}}\label{APP_gradient_ThB}
We first re-express $\mathbf{\bar{H}}_\text{E2}$ from \eqref{Hbar E2} as follows:
\begin{align}\label{Hbar E2 transf}
	\mathbf{\bar{H}}_\text{E2}  \left(\mathbf{\Theta}_\text{B}\right):=&    \begin{bmatrix} \mathbf{c_0} & \mathbf{d} \\
		\mathbf{d}^H  & {0} \end{bmatrix},
\end{align}
where $\mathbf{d}=\mathbf{A}(\mathbf{a}+\mathbf{B}\mathbf{\Theta}_\text{B}\mathbf{c})\in\mathds{C}^{M\times 1}$, $\mathbf{A}=\text{diag} \left( \mathbf{h}^*_\text{$\text{I}_\text{E}$,E} \right) \mathbf{h}^*_\text{A,$\text{I}_\text{E}$} \mathbf{w}^* \mathbf{w}^T$ and $\mathbf{B}=\mathbf{h}^T_\text{A,$\text{I}_\text{B}$}$ are defined below \eqref{EQ_grad_TB}, $\mathbf{a}=\mathbf{h}_{A,E}^T$ and $\mathbf{c}=\mathbf{h}_{I_B,E}^T$ are both column vectors.  Then, $\text{tr}\left(\mathbf{\bar{H}}_\text{E2} \left(\mathbf{\Theta}_\text{B}\right)  \mathbf{\tilde{\Theta}}_\text{E}\right)$ can be rearranged as follows:
\begin{align}
	&\text{tr}\left(\mathbf{\bar{H}}_\text{E2} \left(\mathbf{\Theta}_\text{B}\right)  \mathbf{\tilde{\Theta}}_\text{E}\right)\notag\\
	&=\lambda_2\text{tr}\left(\begin{bmatrix} \mathbf{c_0} & \mathbf{d} \\
		\mathbf{d}^H  & {0} \end{bmatrix}
	\begin{bmatrix} 
		\mid &  & \mid & \mid \\
		\theta_1^{E*}\boldsymbol{\bar{\theta}_\text{E}}& \cdots  & \theta_{N_E}^{E*}\boldsymbol{\bar{\theta}_\text{E}} & \boldsymbol{\bar{\theta}_\text{E}}\\
		\mid &  & \mid & \mid \end{bmatrix}\right)\label{EQ_f1_1}\\
	&=\lambda_2\text{tr}\left(\begin{bmatrix} \mathbf{c_0} & \mathbf{d} \\
		\mathbf{d}^H  & {0} \end{bmatrix}
	\begin{bmatrix} 
		\theta_1^{E*}\boldsymbol{\theta_\text{E}}  &\cdots & \theta_{N_E}^{E*}\boldsymbol{\theta_\text{E}} & \boldsymbol{\theta_\text{E}}\\
		\theta_1^{E*} &  & \theta_{N_E}^{E*} & 1 \end{bmatrix}\right)\label{EQ_f1_2}\\
	&=\lambda_2\left(\left(\sum_{k=1}^{N_E}\theta_k^{E*}\mathbf{c_{0,k}}\right)\boldsymbol{\theta_\text{E}}+\mathbf{d}^T\boldsymbol{\theta_\text{E}}^*+\mathbf{d}^H \boldsymbol{\theta_\text{E}}\right)\label{EQ_f1_3}\\
	&=\lambda_2\left(\tilde{{c}}_0+\boldsymbol{\theta_\text{E}}^H\mathbf{d}+\boldsymbol{\theta_\text{E}}^T\mathbf{d}^*\right),\label{EQ_f1_6}
\end{align}
where \eqref{EQ_f1_1} and \eqref{EQ_f1_2} are by definition of $\mathbf{\tilde{\Theta}}_\text{E}$ and $\boldsymbol{\bar{\theta}_\text{E}}$, respectively. In \eqref{EQ_f1_3}, we define $\mathbf{c_{0,k}}$ as the $k$-th row of $\mathbf{c_{0}}$. In \eqref{EQ_f1_6}, we define the first term inside the bracket in \eqref{EQ_f1_3} as $\tilde{\mathbf{c}}_0$.

Then from \eqref{EQ_f1_6} we have 
\begin{align}
	\nabla_{\boldsymbol{\Theta_\text{B}}} &\text{tr}\left(\mathbf{\bar{H}}_\text{E2} \left(\mathbf{\Theta}_\text{B}\right)  \mathbf{\tilde{\Theta}}_\text{E}\right)=\lambda_2\frac{d}{d \boldsymbol{\Theta_\text{B}}^*}\left(\tilde{{c}}_0+\boldsymbol{\theta_\text{E}}^H\mathbf{d}+\boldsymbol{\theta_\text{E}}^T\mathbf{d}^*\right)\label{EQ_Grad_tr_1}\\
	&=\lambda_2\frac{d}{d \boldsymbol{\Theta_\text{B}}^*}\boldsymbol{\theta}_\text{E}^H\mathbf{d}^*\label{EQ_Grad_tr_2}\\
	&=\lambda_2\frac{d}{d \boldsymbol{\Theta_\text{B}}^*}\boldsymbol{\theta}_\text{E}^H\mathbf{A}^*(\mathbf{a}^*+\mathbf{B}^*\boldsymbol{\Theta_\text{B}}^*\mathbf{c}^*)\label{EQ_Grad_tr_3}\\
	&=\lambda_2\frac{d}{d \boldsymbol{\Theta_\text{B}}^*}\mbox{tr}\left(\mathbf{A}^*(\mathbf{a}^*+\mathbf{B}^*\boldsymbol{\Theta_\text{B}}^*\mathbf{c}^*)\boldsymbol{\theta}_\text{E}^H\right)\label{EQ_Grad_tr_4}\\
	&=\lambda_2\frac{d}{d \boldsymbol{\Theta_\text{B}}^*}\mbox{tr}\left(\mathbf{c}^*\boldsymbol{\theta}_\text{E}^H\mathbf{A}^*\mathbf{B}^*\boldsymbol{\Theta_\text{B}}^*\right)\label{EQ_Grad_tr_5}\\
	&={\lambda_2}\mathbf{B}^H\mathbf{A}^H\boldsymbol{\theta}_\text{E}^*\mathbf{c}^H,\label{EQ_grad_TB12}
\end{align}
where \eqref{EQ_Grad_tr_1} is from \cite[(4.48)]{hjorungnes_2011}, \eqref{EQ_Grad_tr_2} is due to the fact that $\tilde{\mathbf{c}}_0$, $\boldsymbol{\theta}_\text{E}$ and $\mathbf{d}$ do not contain $\boldsymbol{\Theta_\text{B}}^*$, 
\eqref{EQ_Grad_tr_3} is due to the fact that $ \mathbf{r}^T\cdot\mathbf{s}=\mbox{tr}(\mathbf{r}^T\cdot\mathbf{s})=\mbox{tr}(\mathbf{s}\cdot\mathbf{r}^T),$ where $\mathbf{r}$ and $\mathbf{s}$ are column vectors,
\eqref{EQ_Grad_tr_4} is due to the property $\mbox{tr}(\mathbf{P}\mathbf{Q})=\mbox{tr}(\mathbf{Q}\mathbf{P})$ and also $\mathbf{a}^*$ and $\mathbf{A}^*$ are not functions of $\boldsymbol{\Theta_\text{B}}^*$,
\eqref{EQ_grad_TB12} is due to the property $\frac{d}{d\mathbf{Z}^*}\mbox{tr}(\mathbf{P}\mathbf{Z}^*)=\mathbf{P}^T$ \cite[Table 4.3]{hjorungnes_2011}.
The gradient of the second term in $f$ is as follows:
\begin{align}
	&\lambda_2\nabla_{\mathbf{\Theta}_\text{B}}\mathbf{\bar{h}}_\text{E2} \left(\mathbf{\Theta}_\text{B}\right) = \lambda_2\nabla_{\mathbf{\Theta}_\text{B}}\mbox{tr}\left[\mathbf{\bar{h}}_\text{E2} \left(\mathbf{\Theta}_\text{B}\right)\right]\label{EQ_grad_TB2_1}\\
	&\hspace{-0.8 cm}=\frac{\lambda_2}{\sigma_E^2} \nabla_{\mathbf{\Theta}_\text{B}}\mbox{tr}\left[\left( \mathbf{h}_\text{A,E} + \mathbf{h}_\text{$\text{I}_\text{B}$,E} \mathbf{\Theta}_\text{B} \mathbf{h}_\text{A,$\text{I}_\text{B}$} \right)^* \mathbf{Q} \left( \mathbf{h}_\text{A,E} +\mathbf{h}_\text{$\text{I}_\text{B}$,E} \mathbf{\Theta}_\text{B} \mathbf{h}_\text{A,$\text{I}_\text{B}$} \right)^T\right]\label{EQ_grad_TB2_2}\\ 
	&\hspace{-0.8 cm}=\frac{\lambda_2}{\sigma_E^2} \nabla_{\mathbf{\Theta}_\text{B}}\mbox{tr}\left[
	\left( \mathbf{h}_\text{A,E} +\mathbf{h}_\text{$\text{I}_\text{B}$,E} \mathbf{\Theta}_\text{B} \mathbf{h}_\text{A,$\text{I}_\text{B}$} \right)^T
	\left( \mathbf{h}_\text{A,E} + \mathbf{h}_\text{$\text{I}_\text{B}$,E} \mathbf{\Theta}_\text{B} \mathbf{h}_\text{A,$\text{I}_\text{B}$} \right)^* \mathbf{Q}\right]\label{EQ_grad_TB2_3}\\
	&\hspace{-0.8 cm}=\frac{\lambda_2}{\sigma_E^2} \nabla_{\mathbf{\Theta}_\text{B}}\mbox{tr}\left[
	\mathbf{Q} \left( \mathbf{h}_\text{A,E} +\mathbf{h}_\text{$\text{I}_\text{B}$,E} \mathbf{\Theta}_\text{B} \mathbf{h}_\text{A,$\text{I}_\text{B}$} \right)^T
	\left( \mathbf{h}_\text{A,E} + \mathbf{h}_\text{$\text{I}_\text{B}$,E} \mathbf{\Theta}_\text{B} \mathbf{h}_\text{A,$\text{I}_\text{B}$} \right)^* \right]\label{EQ_grad_TB2_4}\\
	=&\frac{\lambda_2}{\sigma_E^2} \nabla_{\mathbf{\Theta}_\text{B}}\mbox{tr}\left[
	\mathbf{c}_1+ \mathbf{c}_2\mathbf{\Theta}_\text{B}^* \right]\label{EQ_grad_TB2_5}\\
	=&\frac{\lambda_2}{\sigma_E^2}\mathbf{c}_2^T,\label{EQ_grad_TB2}
\end{align}
where \eqref{EQ_grad_TB2_1} is due to the fact that $\mathbf{\bar{h}}_\text{E2} \left(\mathbf{\Theta}_\text{B}\right)$ is a scalar, \eqref{EQ_grad_TB2_2} is by definition \eqref{EQ_he2}, and \eqref{EQ_grad_TB2_3} and \eqref{EQ_grad_TB2_4} use tr$(\mathbf{A}\mathbf{B})$=tr$(\mathbf{B}\mathbf{A})$. In \eqref{EQ_grad_TB2_5}, $\mathbf{c}_1$ includes all term not relevant to $\boldsymbol{\Theta_\text{B}}^*$ and $\mathbf{c}_2:=\mathbf{h}_\text{A,$\text{I}_\text{B}$}^*\mathbf{Q}\left( \mathbf{h}_\text{A,E} +\mathbf{h}_\text{$\text{I}_\text{B}$,E} \mathbf{\Theta}_\text{B} \mathbf{h}_\text{A,$\text{I}_\text{B}$} \right)^T\mathbf{h}_\text{$\text{I}_\text{B}$,E}^*$.
After combining \eqref{EQ_grad_TB12} and \eqref{EQ_grad_TB2}, we have the gradient of $f$ with respect to $\mathbf{\Theta}_\text{B}$ as \eqref{EQ_grad_TB}.

\bibliographystyle{IEEEtran}
\bibliography{IEEEabrv,Bibliography}

\end{document}